\documentclass[12pt]{article}
 \input epsf.sty

\begin{document}

\def\ket{\rangle}
\def\bra{\langle}
\def\CA{{\cal A}}
\def\CB{{\cal B}}
\def\CC{{\cal C}}
\def\CD{{\cal D}}
\def\CE{{\cal E}}
\def\CF{{\cal F}}
\def\CG{{\cal G}}
\def\CH{{\cal H}}
\def\CI{{\cal I}}
\def\CJ{{\cal J}}
\def\CK{{\cal K}}
\def\CL{{\cal L}}
\def\CM{{\cal M}}
\def\CN{{\cal N}}
\def\CO{{\cal O}}
\def\CP{{\cal P}}
\def\CQ{{\cal Q}}
\def\CR{{\cal R}}
\def\CS{{\cal S}}
\def\CT{{\cal T}}
\def\CU{{\cal U}}
\def\CV{{\cal V}}
\def\CW{{\cal W}}
\def\CX{{\cal X}}
\def\CY{{\cal Y}}
\def\CZ{{\cal Z}}

\newcommand{\todo}[1]{{\em \small {#1}}\marginpar{$\Longleftarrow$}}
\newcommand{\labell}[1]{\label{#1}}
\newcommand{\bbibitem}[1]{\bibitem{#1}\marginpar{#1}}
\newcommand{\llabel}[1]{\label{#1}\marginpar{#1}}
\newcommand{\dslash}[0]{\slash{\hspace{-0.23cm}}\partial}

\newcommand{\sphere}[0]{{\rm S}^3}
\newcommand{\su}[0]{{\rm SU(2)}}
\newcommand{\so}[0]{{\rm SO(4)}}
\newcommand{\bK}[0]{{\bf K}}
\newcommand{\bL}[0]{{\bf L}}
\newcommand{\bR}[0]{{\bf R}}
\newcommand{\tK}[0]{\tilde{K}}
\newcommand{\tL}[0]{\bar{L}}
\newcommand{\tR}[0]{\tilde{R}}

\newcommand{\btzm}[0]{BTZ$_{\rm M}$}
\newcommand{\ads}[1]{{\rm AdS}_{#1}}
\newcommand{\ds}[1]{{\rm dS}_{#1}}
\newcommand{\eds}[1]{{\rm EdS}_{#1}}
\newcommand{\sph}[1]{{\rm S}^{#1}}
\newcommand{\gn}[0]{G_N}
\newcommand{\SL}[0]{{\rm SL}(2,R)}
\newcommand{\cosm}[0]{R}
\newcommand{\hdim}[0]{\bar{h}}
\newcommand{\bw}[0]{\bar{w}}
\newcommand{\bz}[0]{\bar{z}}
\newcommand{\be}{\begin{equation}}
\newcommand{\ee}{\end{equation}}
\newcommand{\bea}{\begin{eqnarray}}
\newcommand{\eea}{\end{eqnarray}}
\newcommand{\pat}{\partial}
\newcommand{\lp}{\lambda_+}
\newcommand{\bx}{ {\bf x}}
\newcommand{\bk}{{\bf k}}
\newcommand{\bb}{{\bf b}}
\newcommand{\BB}{{\bf B}}
\newcommand{\tp}{\tilde{\phi}}
\hyphenation{Min-kow-ski}

\def\apr{\alpha'}
\def\str{{str}}
\def\lstr{\ell_\str}
\def\gstr{g_\str}
\def\Mstr{M_\str}
\def\lpl{\ell_{pl}}
\def\Mpl{M_{pl}}
\def\varep{\varepsilon}
\def\del{\nabla}
\def\grad{\nabla}
\def\tr{\hbox{tr}}
\def\perp{\bot}
\def\half{\frac{1}{2}}
\def\p{\partial}
\def\perp{\bot}
\def\eps{\epsilon}

\renewcommand{\thepage}{\arabic{page}}
\setcounter{page}{1}

\rightline{hep-th/0206248} \rightline{UPR-1004-T} \vskip 1cm

\centerline{\Large \bf String Interactions in PP-wave  }
\centerline{\Large \bf from ${\cal N} = 4$ Super Yang Mills}
\vskip 0.5 cm
\renewcommand{\thefootnote}{\fnsymbol{footnote}}
\centerline{{ Min-xin Huang \footnote{minxin@sas.upenn.edu} }}
\vskip .5cm \centerline{\it David Rittenhouse Laboratories,
University of Pennsylvania} \centerline{\it Philadelphia, PA
19104, U.S.A.}

\setcounter{footnote}{0}
\renewcommand{\thefootnote}{\arabic{footnote}}

\begin{abstract}
We consider non-planar contributions to the correlation functions
of BMN operators in free ${\cal N} = 4$ super Yang Mills theory.
We recalculate these non-planar contributions from a different
kind of diagram and find some exact agreements. The vertices of
these diagrams are represented by free planar three point
functions, thus our calculations provide some interesting
identities for correlation functions of BMN operators in ${\cal N}
= 4$ super Yang Mills theory. These diagrams look very much like
loop diagrams in a second quantized string field theory, thus
these identities could possibly be interpreted as natural
consequences of the pp-wave/CFT correspondence.\footnote{For
convenience we will call these diagrams "string theory diagram",
although there are reasonable doubts whether these calculations
are truly string theory calculations since it is not known how to
compute general loop amplitudes in a second quantized string
theory. (see a recent paper \cite{taylor} for progress in this
direction.)}

\end{abstract}


\section{Introduction}
\label{intro} The AdS/CFT correspondence states that the ${\cal N}
= 4$ SU(N) Yang-Mills theory  is equivalent to IIB string theory
quantized on the $\ads{5} \times S^5$ background~\cite{jthroat}.
Recently Berenstein, Maldacena and Nastase \cite{BMN} have argued
that IIB superstring theory on a pp-wave background with
Ramond-Ramond flux is dual to a sector of ${\cal{N}}=4$ SU(N)
super Yang-Mills theory containing operators with large R-charge
$J$  . The pp-wave solution of type IIB supergravity has 32
supersymmetries and can be obtained as a Penrose limit of $AdS_5
\times S^5$ \cite{blau}. While the application of the AdS/CFT
correspondence in the usual $AdS \times S$ background is difficult
to go beyond the supergravity approximation on the string theory
side, the string worldsheet theory in the pp-wave background is
exactly solvable, as shown by \cite{metsaev}. More recently, there
have been some progress on the question of string interactions
\cite{seme}-\cite{new}.

It is pointed out in \cite{seme, bn, harvmit} that in BMN limit
some non-planar diagrams of arbitrary genus survive and string
interactions in pp-wave involve two expansion parameters
\begin{equation}
\lambda^{'}=\frac{g_{YM}^2 N}{J^2}=\frac{1}{(\mu p^{+}
\alpha^{'})^2}
\end {equation}

\begin{equation}
g_2=\frac{J^2}{N}=4\pi g_s (\mu p^{+} \alpha^{'})^2
\end{equation}
Here the expansion in $g_2$ comes from non-planar diagrams. There
are operator mixings in this limit. The BMN operators no longer
have well defined conformal dimensions and need to be redefined
order by order \cite{seme, harvmit}. \footnote{It has been argued
earlier that non-planar contributions to large charge operator
correlators are important \cite{Balasubramanian1} \cite{Corley}.
Here in the BMN limit the R-charge goes like $J\sim
N^{\frac{1}{2}}$ and the non-planar diagrams are perturbative in
the expansion parameter $g_2$. If the R-charge is larger,
non-planar diagrams will dominate over planar diagrams
\cite{Balasubramanian1}, and the strings blow up into giant
gravitons by Myers effect \cite{Myers, graviton}. Giant gravitons
are D3 branes described by determinants and subdeterminants
instead of trace operators in CFT \cite{Balasubramanian1} . Open
strings attached to giant gravitons are described in
\cite{Balasubramanian}.} In this paper for simplicity we will only
consider free Yang Mills theory, i.e. we set $\lambda^{'}=0$, so
the only expansion parameter is $g_2$. Also there will be no
anomalous conformal dimensions in this case and we do not need to
consider operator mixing.

It is proposed in \cite{harvmit} that the interaction amplitude
for a single string to split into two strings (or two strings
joining into one string) is related to the three point function of
the corresponding operators in the dual CFT. Using this relation
the authors in \cite{harvmit} are able to compute the second order
correction to the anomalous dimension of the BMN operator from
free planar three point functions and found exact agreement with
computation of the field theory torus contributions to the two
point function. All calculations in \cite{harvmit} were done in
the field theory side but has a clear interpretation from dual
string theory. This proposal has been explicitly checked \cite
{kklp, minxin, chu1, lmp, spvo2} by calculations from light cone
string field theory in pp-wave \cite{spvo1}. \footnote{Another
interesting approach to this question is to use matrix string
theory \cite{matrix}, see \cite{gopakumar}.}

\begin{figure}
  \begin{center}
 \epsfysize=2.0in
   \mbox{\epsfbox{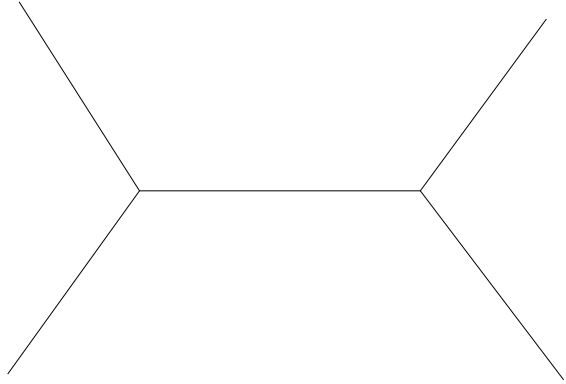}}
\end{center}
\caption{It is recently pointed out in \cite{chu2} that higher
point string interactions in pp-wave can be reduced to cubic
interactions under some double pinching limits. For example, the
skeleton diagram with $s$, $t$ and $u$ channels appear in the
computation of a planar four point function
$\langle\bar{O}_1\bar{O}_2O_3O_4\rangle$ as we take some specific
double pinching limits. We will only need to consider cubic
interactions in our calculations of the string theory diagram.}
\label{F1}
\end{figure}

It has been pointed out that in type IIB light cone string field
theory string interactions should contain quartic or higher order
contact interactions in addition to cubic interactions \cite{GS},
but for some unknown reasons in pp-wave we only need to consider
cubic vertex, representing string joining and splitting. This is
justified by our calculations where we find precise agreements by
only including cubic vertices. Similar point of view is also taken
in a recent paper \cite{chu2} (see figure \ref{F1}). It would be
interesting if higher order interactions indeed vanish in pp-wave
light cone string field theory and we leave it for future works.
In this paper we will represent the vertices in string theory
diagrams with free planar three point functions according to the
proposal of \cite{harvmit}.\footnote{The matrix element in
\cite{harvmit} contain a prefactor and a vertex, but for some
unknown reasons we do not need to use the prefactor. We leave the
explanation to future works.} Some free planar three point
functions involving BMN operators have been computed in
\cite{harvmit}. Specifically, we have (Assuming $m\neq 0$ and
$n\neq 0$)

\begin {equation} \label{planar1}
\langle\bar{O}^JO^{J_1}O^{J_2}\rangle=\frac{g_2}{\sqrt{J}}\sqrt{x(1-x)}
\end {equation}

\begin {equation} \label{planar2}
\langle\bar{O}^J_{0}O^{J_1}O^{J_2}_{0}\rangle=\frac{g_2}{\sqrt{J}}x^{\frac{1}{2}}(1-x)
\end {equation}

\begin {equation} \label{planar3}
\langle\bar{O}^J_{00}O^{J_1}_{0}O^{J_2}_{0}\rangle=\frac{g_2}{\sqrt{J}}x(1-x)
\end {equation}

\begin {equation} \label{planar4}
\langle\bar{O}^J_{m,-m
}O^{J_1}_{0}O^{J_2}_{0}\rangle=-\frac{g_2}{\sqrt{J}}\frac{\sin^2(\pi
mx)}{\pi^2m^2}
\end {equation}

\begin{equation} \label{planar5}
\langle\bar{O}^J_{00}O^{J_1}_{00}O^{J_2}\rangle=\frac{g_2}{\sqrt{J}}x^{\frac{3}{2}}\sqrt{1-x}
\end{equation}

\begin {equation} \label{planar6}
\langle\bar{O}^J_{m,-m}O^{J_1}_{n,-n}O^{J_2}\rangle=\frac{g_2}{\sqrt{J}}x^{\frac{3}{2}}\sqrt{1-x}\frac{\sin^2(\pi
mx)}{\pi^2 (mx-n)^2}
\end {equation}

\begin {equation} \label{planar7}
\langle\bar{O}^J_{00}O^{J_1}_{n,-n}O^{J_2}\rangle=0
\end {equation}
where $x=J_1/J$ and $J=J_1+J_2$. Note the spacetime dependences of
two point and three point functions in conformal field theory are
determined by conformal symmetry. Here and elsewhere in this paper
we have omitted the factors of spacetime dependence in the
correlators. The definition of the properly normalized chiral and
BMN operators are

\begin{equation}
O^{J}=\frac{1}{\sqrt{N^JJ}}TrZ^J
\end{equation}

\begin {equation}
O^{J_1}_{0}=\frac{1}{\sqrt{N^{J_1+1}}} Tr(\phi^{I_1} Z^{J_1})
\end {equation}

\begin {equation}
O^{J_2}_{0}=\frac{1}{\sqrt{N^{J_2+1}}} Tr(\phi^{I_2} Z^{J_2})
\end {equation}

\begin{equation}\label{bnnop}
O^J_{m,-m} = \frac1{\sqrt{JN^{J+2}}} \sum_{l=0}^Je^{2\pi iml/J}
Tr(\phi^{I_1} Z^l\phi^{I_2} Z^{J-l}).
\end{equation}
Here $\phi^{I_1}$ and $\phi^{I_2}$ represent excitations in two of
the eight transverse directions.

The proposal of this paper is that we can calculate non-planar
contributions to BMN correlation functions in free Yang Mills
theory from string theory point of view. The details of how to do
the calculation will be clear from our specific examples. Some
non-planar contributions to the two point and three point
functions of BMN operators have been computed on the field theory
side in \cite {seme, harvmit}. For example, the free torus (genus
one) two functions of BMN operators are

\begin{eqnarray}\label{torus}
&&\langle \bar{O}_{n,-n}^J O_{m,-m}^J \rangle_{torus}   \\
&& = \frac{g_2^2}{24}, \qquad\qquad\qquad\qquad\qquad\qquad\qquad\qquad\qquad\qquad\qquad\qquad m=n=0; \nonumber\\
&& = 0,  \qquad\qquad\qquad\qquad\qquad\qquad\qquad\qquad\qquad\qquad m=0, n\neq0~~or~~n=0, m\neq0; \nonumber\\
&& =g_2^2(\frac{1}{60} - \frac{1}{24 \pi^2 m^2} + \frac{7}{16 \pi^4 m^4}), \qquad\qquad\qquad\qquad\qquad\qquad\qquad m=n\neq0; \nonumber\\
&& =\frac{g_2^2}{16\pi^2m^2} ( \frac{1}{3}+\frac{35}{8\pi^2m^2}),
\qquad\qquad\qquad\qquad\qquad\qquad\qquad\qquad m=-n\neq0; \nonumber\\
&& =\frac{g_2^2}{4\pi ^{2}(m-n)^2} ( \frac{1}{3}+\frac{1}{\pi
^2n^2}+\frac{1}{\pi ^2m^2}-\frac{3}{2\pi ^2mn}-\frac{1}{2\pi
^2(m-n)^2}), ~~all~other~cases \nonumber
\end{eqnarray}
The paper is organized as follows. In section \ref{twopoint1} we
reproduce equation (\ref{torus}) from one loop string propagation
diagram calculation. In section \ref{cubic} we calculate the torus
contribution to a three point function involving BMN operators
both in field theory side and in string theory side. We also find
nontrivial agreements in this case. In section \ref{twopoint2} and
appendix \ref{cancellation} we do more calculations giving more
evidences of our proposal.

\section{One loop string propagation}

\begin{figure}
  \begin{center}
 \epsfysize=2.5in
   \mbox{\epsfbox{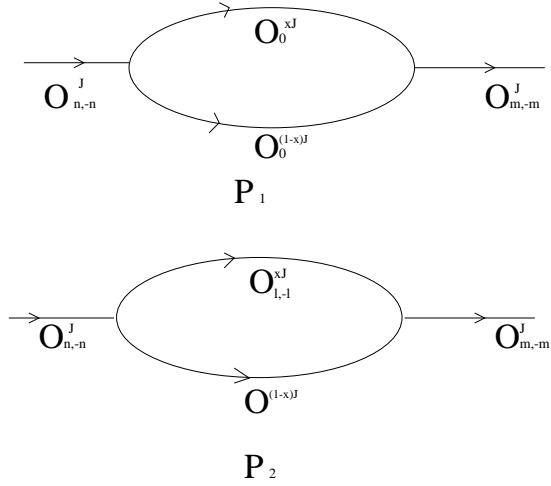}}
\end{center}
\caption{There are 2 diagrams contributing the one loop string
propagation. The BMN string $O^J_{n,-n}$ can split into two
strings $O^{J_1}_{l,-l}$, $O^{J_2}$ or $O^{J_1}_0$, $O^{J_2}_0$
and joining back into another string $O^J_{m,-m}$. We denote
contributions to these two diagrams $P_1$ and $P_2$.} \label{F2}
\end{figure}

\label{twopoint1} We consider a single string propagating in the
pp-wave background. We expect the one loop correction to the
string propagation to be the torus contribution to the two point
function of corresponding BMN operators. On the other hand, the
one loop amplitude can be calculated by summing over the
amplitudes of the string splitting into two strings and then
joining back into a single string. The cubic vertices of string
splitting and joining can be represented by free planar three
point functions.  There are two diagrams associated with this
process as shown in figure \ref{F2}. The BMN string $O^J_{n,-n}$
can split into two strings $O^{J_1}_{l,-l}$, $O^{J_2}$ or
$O^{J_1}_0$, $O^{J_2}_0$ and joining back into another string
$O^J_{m,-m}$. We denote the contributions from these two processes
by $P_1$ and $P_2$. Then

\begin{eqnarray}\label{P1}
P_1&=&\sum_{J_1=0}^{J}\langle \bar{O}^J_{n,-n} O^{J_1}_0 O^{J_2}_0
\rangle_{planar} \langle \bar{O}^{J_1}_0 \bar{O}^{J_2}_0
O^J_{m,-m} \rangle_{planar}\\&&\nonumber
=g_2^2\int_0^1dx\frac{\sin^2(m\pi x)}{m^2\pi^2}\frac{\sin^2(n\pi
x)}{n^2\pi^2}
\end{eqnarray}

\begin{eqnarray}\label{P2}
P_2&=&\sum_{J_1=0}^{J}\sum_{l=-\infty}^{+\infty} \langle
\bar{O}^J_{n,-n} O^{J_1}_{l,-l} O^{J_2} \rangle_{planar} \langle
\bar{O}^{J_1}_{l,-l} \bar{O}^{J_2} O^J_{m,-m}
\rangle_{planar}\\&&\nonumber =g_2^2\sum_{l=-\infty}^{+\infty}
\int_0^1dxx^{3}(1-x)\frac{\sin^2(m\pi
x)}{\pi^2(mx-l)^2}\frac{\sin^2(n\pi x)}{\pi^2(nx-l)^2}
\end{eqnarray}

\begin{figure}
  \begin{center}
 \epsfysize=2.5in
   \mbox{\epsfbox{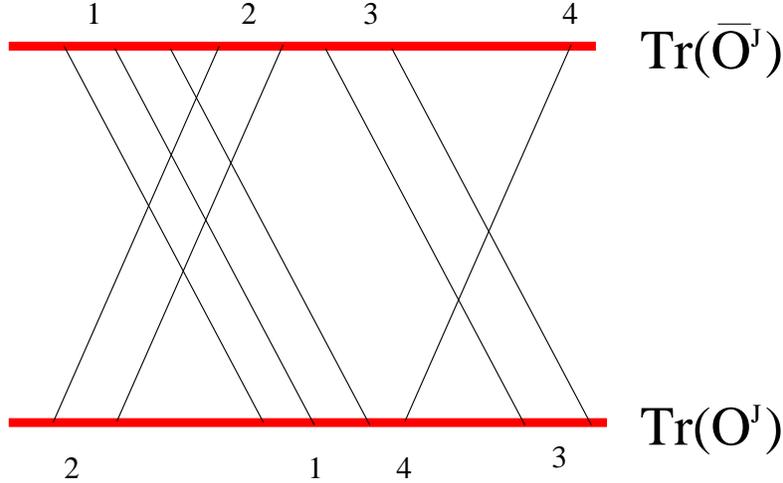}}
\end{center}
\caption{Feymann diagram of torus contraction of large $N$ gauge
indices \cite{seme,harvmit}. We are contracting non-planarly by
dividing the string into 4 segments.} \label{F3}
\end{figure}

The string theory diagrams are computed by multiplying all
vertices and summing over all possible intermediate operators.
Here we do not need to use propagators in calculating the
diagrams. In large $J$ limit we can approximate the sum in $J_1$
by a integral $\sum_{J_1=0}^{J}=J\int^1_0 dx$. It is
straightforward to put equations (\ref{planar1}) to
(\ref{planar7}) into equations (\ref{P1}) (\ref{P2}) and
explicitly compute the sum and integral. We find an agreement with
equation (\ref{torus}) in all 5 cases

\begin{equation} \label{a1}
\langle \bar{O}^J_{n,-n} O^J_{m,-m}
\rangle_{torus}=\frac{1}{2}(P_1+P_2)
\end{equation}
Here the $\frac{1}{2}$ can be thought of as the symmetry factor of
the string theory diagrams. The symmetry factor can be understood
from the example of free torus two point function of chiral
operators, which is computed in the field theory side as shown in
figure \ref{F3} \cite{seme, harvmit}. The twistings in the large
$N$ gauge index contractions can be thought of intuitively as
string splitting and rejoining. In appendix \ref{count} we give an
argument why we have overcounted by a factor of $2$ when we do
string theory diagrams. In more general cases of one loop cubic
interaction and two loop propagation diagrams the symmetry factors
will be determined by more complicated combinatorics and will
generally differ from the symmetry factors in usual Feymann
diagrams in quantum field theory. In appendix \ref{count} we
derive  the symmetry factors of one loop cubic diagrams and find
agreements with direct calculations of field theory and string
theory diagrams. We have not determined the symmetry factors for
two loop propagation diagrams. Nevertheless we can still do some
interesting calculations in section \ref{twopoint2} and appendix
\ref{cancellation} without knowing the symmetry factors.

One can also easily calculate the one loop string propagation
diagram for chiral operators $O^J$ and $O^J_0$. In both cases
there is only one diagram. The results are again agree with the
field theory calculations by the symmetry factor of $\frac{1}{2}$.

\section{One loop cubic string interaction }
\label{cubic}
\subsection{Free torus three point functions of BMN operators}
\begin{figure}
  \begin{center}
 \epsfysize=2.5in
   \mbox{\epsfbox{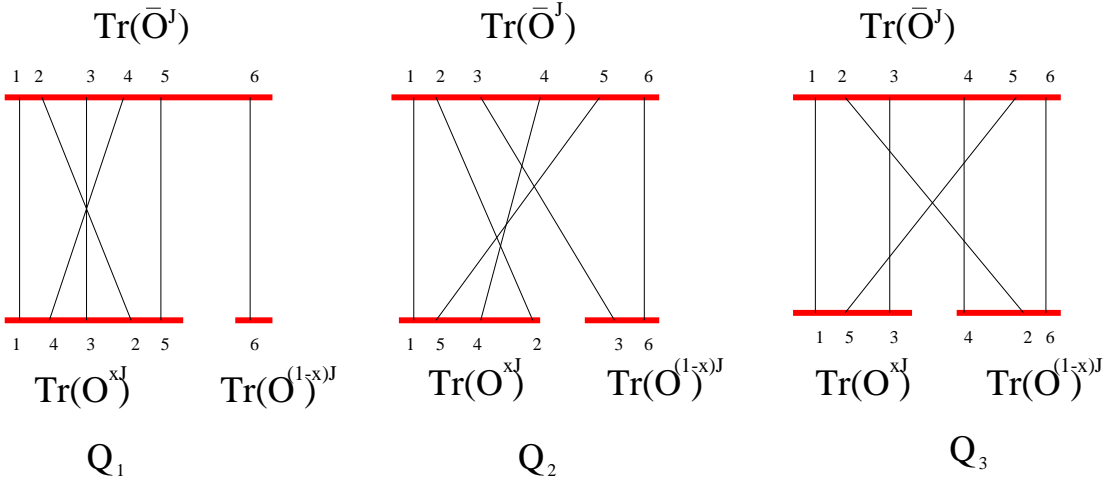}}
\end{center}
\caption{There are three diagrams contribute to the torus three
point function. We denote their contributions by $Q_1$, $Q_2$ and
$Q_3$. Here we use single line notation. One would check these
diagrams indeed have a power of $1/N^3$ in double line notation.
The top line and the bottom lines represent the long string and
short strings. In the three diagrams we have divided the long
string into 6 groups and the short strings into (5,1), (4,2) and
(3,3) groups. Each group is represented by a single line here.}
\label{F4}
\end{figure}

The non-planar contributions to the three point functions of large
charge chiral operators have been computed to arbitrary genus
using Gaussian matrix model \cite{seme}. Here we calculate the
torus three point functions using gauge theory Feymann diagram and
generalize calculations to BMN operators. The calculations in this
subsection follow very closely as in \cite{seme, harvmit}. First
we consider three point function of chiral operators
$\langle\bar{O}^JO^{J_1}O^{J_2}\rangle$. There are 3 types of
torus diagrams as shown in Figure \ref{F4}. We denote the
contributions from these 3 diagrams $Q_1$, $Q_2$ and $Q_3$. We can
see $Q_1$ is to divide one of the small operators into 5 groups,
so we have a factor of $\frac{1}{4!}$ . $Q_2$ is to divide one of
the small operators into 4 groups and the other one into 2 groups,
so we have a factor of $\frac{1}{3!1!}$. $Q_3$ is to divide both
small operators into 3 group, so the factor is $\frac{1}{2!2!}$.
We caution the reader here we have overcounted by a factor of $2$
in $Q_2$ and a factor of $3$ in $Q_3$ by cyclicity. The final
answer is
\begin{equation}
Q_1=\frac{1}{24}\frac{g_{2}^{3}}{\sqrt{J}}\sqrt{x(1-x)}[x^4+(1-x)^4]
\end{equation}
\begin{equation}
Q_2=\frac{1}{12}\frac{g_{2}^{3}}{\sqrt{J}}\sqrt{x(1-x)}[x^3(1-x)+x(1-x)^3]
\end{equation}
\begin{equation}
Q_3=\frac{1}{12}\frac{g_{2}^{3}}{\sqrt{J}}\sqrt{x(1-x)}[x^2(1-x)^2]
\end{equation}
Here again $x=J_1/J$. One can easily check this calculation by
expand to the first order the three point function equation (3.4)
in \cite{seme}.

Now we consider the three point function involving BMN operator
$\langle\bar{O}^J_{m,-m}O^{J_1}_{0}O^{J_2}_{0}\rangle$. The
calculation is to insert two scalars in the diagrams in Figure
\ref{F4} and sum over all positions with phases \cite{seme,
harvmit}. We denote contributions from the 3 diagrams $Q_1^{'}$,
$Q_2^{'}$ and $Q_3^{'}$, then
\begin{eqnarray}
Q_1^{'}&=&\frac{g_2^3}{\sqrt{J}}\int_0^{1}dj_1dj_2dj_3dj_4dj_5dj_6\delta(j1+j2+j3+j4+j5-x)\delta(j_6-(1-x))
\nonumber \\&& \int_0^{x}dy_1 e^{2\pi imy_1} \int_x^{1}dy_2
e^{-2\pi imy_2} + (x\rightarrow(1-x))
\end{eqnarray}
\begin{eqnarray}
Q_2^{'}&=&\frac{1}{2}\frac{g_2^3}{\sqrt{J}}\int_0^{1}dj_1dj_2dj_3dj_4dj_5dj_6\delta(j1+j2+j4+j5-x)\delta(j_3+j_6-(1-x))
\nonumber  \\&& \nonumber
(\int_0^{j_1+j_2}+\int_{j_1+j_2+J_3}^{j_1+j_2+j_3+j_4+j_5})
e^{2\pi imy_1}dy_1
(\int_{j_1+j_2}^{j_1+j_2+j_3}+\int_{j_1+j_2+j_3+j_4+j_5}^{1})e^{-2\pi
imy_2}dy_2 \\&& + (x\rightarrow(1-x))
\end{eqnarray}
\begin{eqnarray}
Q_3^{'}&=&\frac{1}{3}\frac{g_2^3}{\sqrt{J}}\int_0^{1}dj_1dj_2dj_3dj_4dj_5dj_6\delta(j1+j3+j5-x)\delta(j_2+j_4+j_6-(1-x))
 \nonumber \\&& \nonumber
(\int_0^{j_1}+\int_{j_1+j_2}^{j_1+j_2+j_3}+\int_{j_1+j_2+j_3+j_4}^{j_1+j_2+j_3+j_4+j_5})
e^{2\pi imy_1}dy_1 \\&&
(\int_{j_1}^{j_1+j_2}+\int_{j_1+j_2+j_3}^{j_1+j_2+j_3+j_4}+\int_{j_1+j_2+j_3+j_4+j_5}^{1})e^{-2\pi
imy_2}dy_2
\end{eqnarray}

Calculations of these integrals give
\begin{eqnarray}\label{e1}
Q_1^{'}=\frac{g_2^3}{\sqrt{J}}\frac{1}{24}(-\frac{\sin^2(m\pi
x)}{m^2\pi^2})(x^4+(1-x)^4)
\end{eqnarray}
\begin{eqnarray}
Q_2^{'}&=&\frac{g_2^3}{\sqrt{J}}\frac{1}{24m^6\pi^6}
[3-3m^2\pi^2x(1-x)-2m^4\pi^4(1-x)x^3\nonumber \\&&
+(-3-3x(1-x)m^2\pi^2)\cos(2m\pi x)+(3(1-2x)m\pi-m^3\pi^3
x^3)\sin(2m\pi x)]\nonumber\\&&+ (x\rightarrow(1-x))\label{e2}
\end{eqnarray}
\begin{eqnarray}
Q_3^{'}&=&\frac{g_2^3}{\sqrt{J}}\frac{1}{16m^6\pi^6}
[-(3+(-1-2x+2x^2)m^2\pi^2+2x^2(1-x)^2m^4\pi^4)\nonumber\\&&+(3-(1-2x)^2m^2\pi^2)\cos(2m\pi
x)-3(1-2x)m\pi\sin(2m\pi x)]\label{e3}
\end{eqnarray}

\subsection{String theory loop diagram calculations}
\begin{figure}
  \begin{center}
 \epsfysize=3.0in
   \mbox{\epsfbox{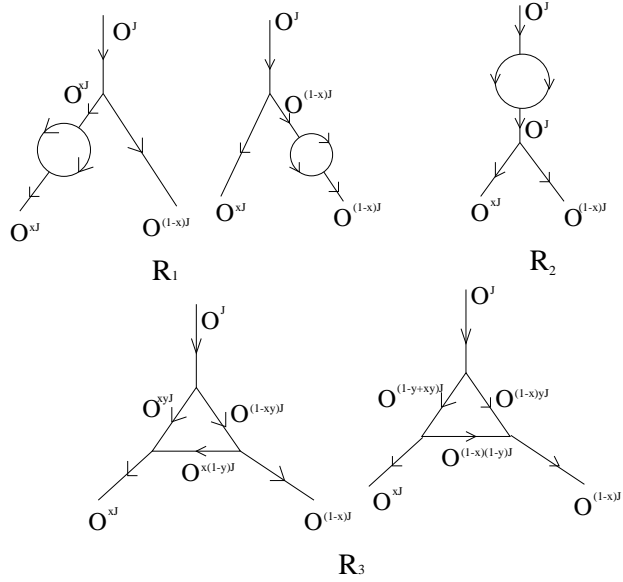}}
\end{center}
\caption{String theory diagrams contribute to
$\langle\bar{O}^JO^{J_1}O^{J_2}\rangle_{torus}$ organized in 3
groups. Diagrams in $R_1$ and $R_2$ are corrections to one
particle propagator while diagrams $R_3$ are amputated. The two
diagrams in $R_3$ are symmetric by exchange of the two decayed
operators.} \label{F5}
\end{figure}

First we consider the simple case of one loop diagrams of chiral
operators $\langle\bar{O}^JO^{J_1}O^{J_2}\rangle_{torus}$. The
diagrams are depicted in figure \ref{F5}. We classify the diagrams
into three groups and denote their contributions $R_1$, $R_2$ and
$R_3$. $R_1$ and $R_2$ are propagator corrections to planar three
point functions. It is obvious that
\begin{equation}
R_1=\frac{g_2^3}{\sqrt{J}}\frac{1}{12}\sqrt{x(1-x)}[x^4+(1-x)^4]
\end{equation}
\begin{eqnarray}
R_2=\frac{g_2^3}{\sqrt{J}}\frac{1}{12}\sqrt{x(1-x)}
\end{eqnarray}
Notice the sum over operators in the fist diagram of $R_3$ gives a
integral $Jx\int_0^1 dy$. The vertices in the first diagram of
$R_3$ are
\begin {equation}
\langle\bar{O}^JO^{xyJ}O^{(1-xy)J}\rangle=\frac{g_2}{\sqrt{J}}\sqrt{xy(1-xy)}
\end {equation}
\begin {equation}
\langle\bar{O}^{xyJ}\bar{O}^{x(1-y)J}O^{xJ}\rangle=\frac{g_2}{\sqrt{J}}x^{\frac{3}{2}}\sqrt{y(1-y)}
\end {equation}
\begin {equation}
\langle\bar{O}^{(1-xy)J}O^{x(1-y)J}O^{(1-x)J}\rangle
=\frac{g_2}{\sqrt{J}}(1-xy)^{\frac{3}{2}}(\frac{1-x}{1-xy})^{\frac{1}{2}}(\frac{x(1-y)}{1-xy})^{\frac{1}{2}}
\end {equation}
So we find
\begin{eqnarray}
R_3&&=\int_{0}^{1}Jxdy(\frac{g_2}{\sqrt{J}}\sqrt{xy(1-xy)})
(\frac{g_2}{\sqrt{J}}x^{\frac{3}{2}}\sqrt{y(1-y)})
(\frac{g_2}{\sqrt{J}}(1-xy)^{\frac{3}{2}}
(\frac{1-x}{1-xy})^{\frac{1}{2}}(\frac{x(1-y)}{1-xy})^{\frac{1}{2}})\nonumber
\\&&
+ (x\rightarrow(1-x)) \nonumber
\\&& =\frac{g_2^3}{\sqrt{J}}\frac{1}{12}\sqrt{x(1-x)}(x^3+(1-x)^3+x^3(1-x)+x(1-x)^3)
\end{eqnarray}
Now we can  write $R_1$, $R_2$ and $R_3$ in terms of $Q_1$, $Q_2$
and $Q_3$. We find
\begin{eqnarray}
&& R_1=2Q_1 \nonumber \\&& R_2=2Q_1+4Q_2+6Q_3  \nonumber \\&&
R_3=2Q_1+2Q_2 \label{factor}
\end{eqnarray}
This is in agreement with equation (\ref{factor1}). The total
contribution to torus three point function is $Q_1+Q_2+Q_3$. In
terms of $R_1$, $R_2$ and $R_3$ it is
\begin{equation} \label{correspondence}
Q_1+Q_2+Q_3=\frac{1}{6}(R_1+R_2+R_3)
\end{equation}
Thus the symmetry factors of all diagrams in figure \ref{F5} are
$\frac{1}{6}$.

\begin{figure}
  \begin{center}
 \epsfysize=2.5in
   \mbox{\epsfbox{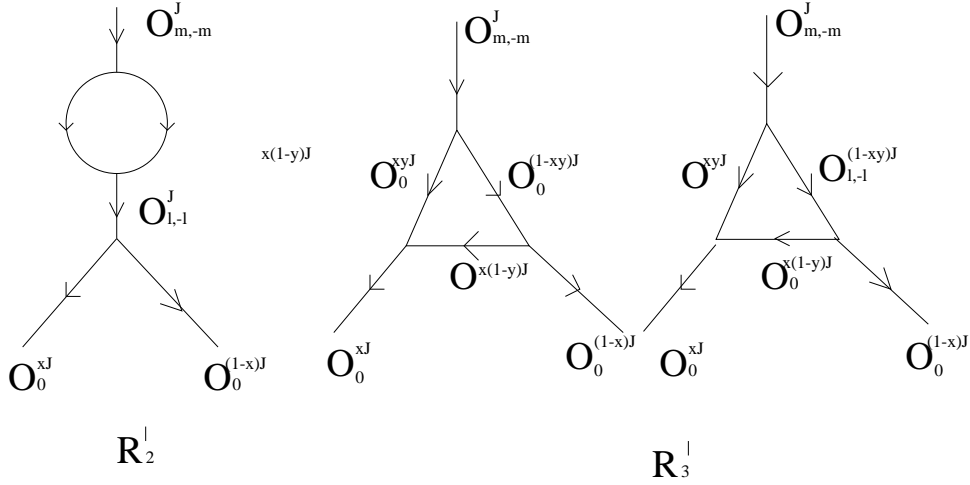}}
\end{center}
\caption{String theory diagrams contribute to
$\langle\bar{O}^J_{m,-m}O^{J_1}_0O^{J_2}_0\rangle_{torus}$.
Diagrams can be organized into 3 groups similar to those in figure
\ref{F5}. Here we only draw two groups which are different from
figure \ref{F5}. There are 4 diagrams in $R_3^{'}$. We only draw 2
of them. The other 2 diagrams is symmetric to what we draw by
exchange of the 2 decayed operators.} \label{F6}
\end{figure}

Now we condiser three point function with a BMN operator
$\langle\bar{O}^J_{m,-m}O^{J_1}_0O^{J_2}_0\rangle_{torus}$. String
theory diagrams contributing to this process are depicted in
figure \ref{F6}. Again we classify the diagrams by 3 groups and
denote their contributions by $R_1^{'}$, $R_2^{'}$ and $R_3^{'}$.
The calculation of $R_1^{'}$ is the same as before. But in the
case of $R_2^{'}$, we need to sum over all possible operators that
is related to $O^J_{m,-m}$ by one loop propagation. The summation
can be done by the summation formulae in appendix \ref{summation}.
\begin{equation}
R_1^{'}=\frac{g_2^3}{\sqrt{J}}\frac{1}{12}(-\frac{\sin^2(m\pi
x)}{m^2\pi^2})[x^4+(1-x)^4]\label{e4}
\end{equation}
\begin{eqnarray}
R_2^{'}&&=\sum_{l=-\infty}^{+\infty}2\langle\bar{O}^J_{m,-m}O^{J}_{l,-l}\rangle_{torus}
\langle\bar{O}^J_{l,-l}O^{J_1}_{0}O^{J_2}_{0}\rangle_{planar}
\nonumber \\&& =\frac{g_2^3}{\sqrt{J}}\frac{1}{24m^6\pi^6}
[-3-(1+2x-2x^2)^2m^4\pi^4+3(3-2x+2x^2)m^2\pi^2 \nonumber \\&&
+(3-3(3-4x+4x^2)m^2\pi^2+(1-4x+6x^2-4x^3+2x^4)m^4\pi^4)\cos(2m\pi
x)\nonumber \\&&
-(3(1-2x)m\pi+4(-1+3x-3x^2+2x^3)m^3\pi^3)\sin(2m\pi x)]\label{e5}
\end{eqnarray}
The calculation of $R_3^{'}$ involves two diagrams and their
symmetric partners by exchanging of the two decayed operators. we
also need to use the summation formulae in appendix
\ref{summation}. The result of doing sum and integrals is
\begin{eqnarray}
R_3^{'}&=&\frac{g_2^3}{\sqrt{J}}\frac{1}{24m^6\pi^6}
[12+12x(x-1)m^2\pi^2+(-1+6x^2-12x^3+6x^4)m^4\pi^4 \nonumber
\\&&+(-12+12x(x-1)m^2\pi^2+(1-4x+6x^2-4x^3+2x^4)m^4\pi^4)\cos(2m\pi
x) \nonumber
\\&&+(12(1-2x)m\pi+2(1-3x+3x^2-2x^3)m^3\pi^3)\sin(2m\pi
x)]\label{e6}
\end{eqnarray}
Using equation (\ref{e1}), (\ref{e2}), (\ref{e3}), (\ref{e4}),
(\ref{e5}) and (\ref{e6}) one readily check
\begin{eqnarray}
&& R_1^{'}=2Q_1^{'} \nonumber \\&&
R_2^{'}=2Q_1^{'}+4Q_2^{'}+6Q_3^{'} \nonumber
\\&& R_3^{'}=2Q_1^{'}+2Q_2^{'} \label{factor2}
\end{eqnarray}
And the total contribution to the torus three point function is
\begin{equation} \label{correspondence1}
\langle\bar{O}^J_{m,-m}O^{J_1}_{0}O^{J_2}_{0}\rangle_{torus}=Q_1^{'}+Q_2^{'}+Q_3^{'}=\frac{1}{6}(R_1^{'}+R_2^{'}+R_3^{'})
\end{equation}
Thus we have found the agreement between field theory and string
theory calculations. It would be also interesting to do
calculations in the case of three point functions involving two
BMN operators such as
$\langle\bar{O}^J_{m,-m}O^{J_1}_{n,-n}O^{J_2}\rangle_{torus}$.

\section{Two loop string propagation}
\label{twopoint2}
\begin{figure}
  \begin{center}
 \epsfysize=3.0in
   \mbox{\epsfbox{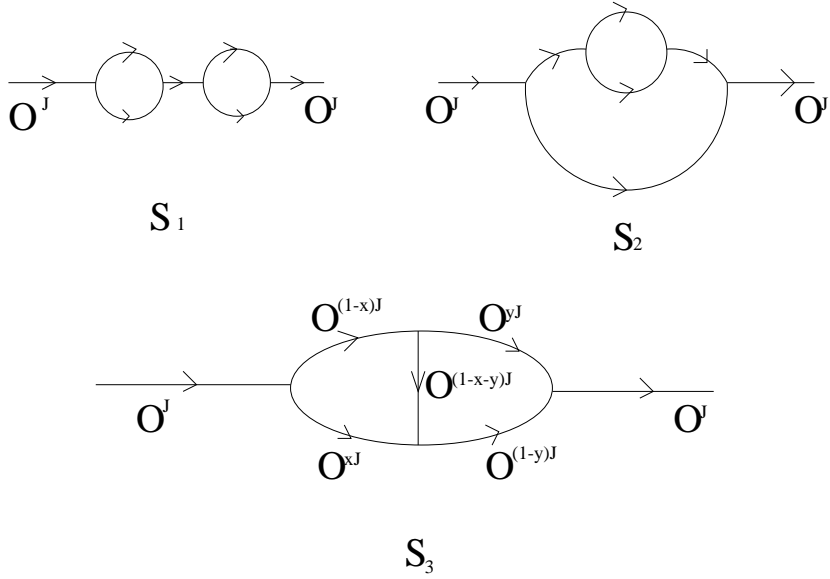}}
\end{center}
\caption{There are three diagrams contribute to
$\langle\bar{O}^JO^{J}\rangle_{genus~2}$. $S_1$ and $S_2$ are
unamputated diagrams while $S_3$ is a amputated diagram. }
\label{F7}
\end{figure}

First we consider the vacuum operator
$\langle\bar{O}^JO^{J}\rangle_{genus~2}$. The three diagrams
$S_1$, $S_2$ and $S_3$ are depicted in figure \ref{F7}. $S_1$ and
$S_2$ are directly related to torus two point function. $S_3$ is a
integral in the range of $x+y<1$ as shown in the diagram. We
calculate these diagrams
\begin{equation}
S_1=\frac{1}{144}g_2^4
\end{equation}
\begin{eqnarray}
S_2&=&g_2^4\int_{0}^{1}dx~x(1-x)\frac{1}{12}x^4 \nonumber
\\&&=\frac{1}{504}g_2^4
\end{eqnarray}
\begin{eqnarray}
S_3&=&g_2^4\int_{0<x,y,x+y<1}dxdy~x(1-x)y(1-y)(1-x-y) \nonumber
\\&&=\frac{1}{280}g_2^4
\end{eqnarray}
Suppose the symmetry factors of the three diagrams $S_1$, $S_2$
and $S_3$ are $a_1$, $a_2$ and $a_3$, then from genus 2 two point
function results in \cite{seme, harvmit} we will require
\begin{equation}
\frac{a_1}{144}+\frac{a_2}{504}+\frac{a_3}{280}=\frac{1}{5!2^4}
\end{equation}

\begin{figure}
  \begin{center}
 \epsfysize=2.5in
   \mbox{\epsfbox{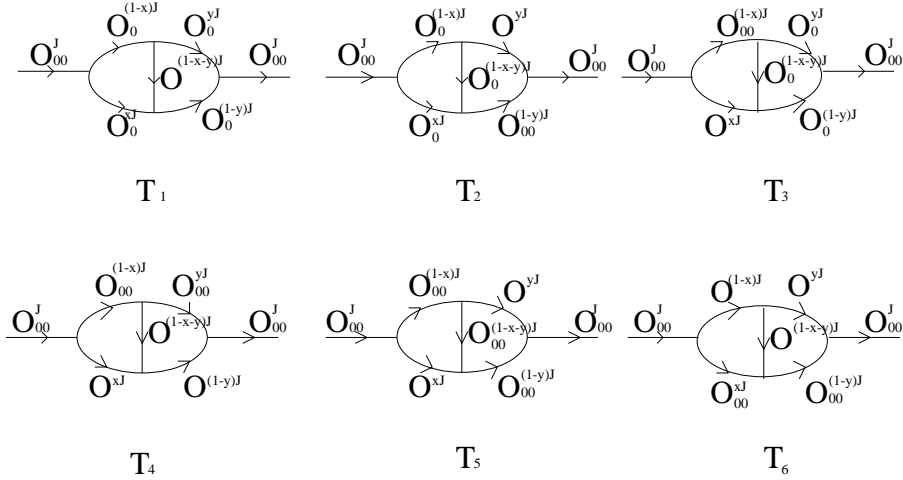}}
\end{center}
\caption{String theory diagrams of
$\langle\bar{O}^J_{00}O^{J}_{00}\rangle_{genus~2}$. We only draw
amputated diagrams since unamputated diagrams are the same as in
figure \ref{F7}. Notice diagrams $T_2$ and $T_3$, $T_4$ and $T_6$
are equal by symmetry. Also diagrams $T_1$, $T_2$ and $T_3$ need
to be multiplied by a factor of 2 since there are two ways to
separate the two different scalar insertions. Interestingly, the
sum of all these 6 diagrams give the same result as a single
diagram $S_3$ in the figure \ref{F7}, as expected from analytic
results of \cite{seme, harvmit}. } \label{F8}
\end{figure}

The genus 2 two point function of BMN operators are computed in
\cite{harvmit}. It would be interesting to determine the symmetry
factors by combinatorics argument as in appendix \ref{count} or by
analytic calculations of string theory diagrams and comparing with
equation (C.36) in \cite{harvmit}. The string theory calculation
of $\langle\bar{O}^J_{m,-m}O^{J}_{n,-n}\rangle_{genus~2}$ will
involve 6 amputated diagrams. In cases of $m=n=0$ or $m=0$, $n\neq
0$ the diagrams are easy to calculate. For $m=n=0$ the sum of the
six diagrams in figure \ref{F8}  interestingly gives the same
result as a single diagram $S_3$. The readers can easily check
(note diagrams $T_1$, $T_2$ and $T_3$ have double contributions)
\begin{equation}
2T_1+2T_2+2T_3+T_4+T_5+T_6=S_3
\end{equation}
This is expected since higher genus contributions to the
correlators of chiral operators are the same regardless of the
number of supergravity excitations \cite{seme, harvmit}. In the
appendix \ref{cancellation} we compute the case $m=0$,$n\neq 0$
and find all diagrams cancel, as expected from the genus 2 two
point function equation (C.36) in \cite{harvmit}.

\section{Conclusion}
In this paper we have described string interaction in pp-wave from
${\cal N} = 4$ super Yang Mills. From our proposal we can
effectively compute the free field correlation functions of BMN
operators to arbitrary genus by diagrammatic expansion in terms of
free planar three point functions. It would be interesting to
perform more detailed calculations as pointed out in the text or
try to give a general analytic proof of these mathematical
identities at arbitrary genus. It is also interesting to
generalize the results to interacting field theory. In this case
the Yang Mills theory is perturbative in two parameters
$\lambda^{'}$ and $g_2$ and there are complications of operator
mixing and redefining in this double scaling limit \cite{seme,
harvmit}. One will need to represent the cubic string vertex with
interacting planar three point functions instead of the free
planar three point functions we used. The planar three point
functions in the first order expansion of $\lambda^{'}$ have been
computed in \cite{chu1}. It would be interesting to use their
results to do the calculations and compare with non-planar
contributions to correlators in interacting Yang Mills theory. But
we should caution the readers the relation between three point
vertex in field theory and string theory is still unclear at
nonzero $\lambda^{'}$ \cite{new}. The three point vertex in
pp-wave light cone string field theory is a very complicated
smoothly interpolating function which involves fractional powers
in small $\lambda^{'}$ expansion \cite{spvo2, new}.

The natural question to ask is what we are really doing here. Can
we interpret the discovered identities as consequences of
pp-wave/CFT correspondence, or are they just mathematical
coincidence and possibly related to some unknown properties of
field theory itself? The answer is not convincingly clear at this
point although we have been inclined to the former explanation.
One possible interpretation is wave function renormalization.
While the authors in \cite{harvmit} computed energy correction,
what we did in section \ref{twopoint1} looks like wave function
renormalization in quantum mechanics.\footnote{We thank Asad Naqvi
for pointing out this to us.} Remember in quantum perturbation
theory the first order wave function renormalization is
\begin{equation}
\langle n|n\rangle=1+\sum_{k\neq n}\frac{|V_{kn}|^2}{(E_n-E_k)^2}
\end{equation}
In string field theory the matrix elements contain a prefactor
which exactly cancels $(E_n-E_k)^2$ in the denominator
\cite{harvmit}. That is why we never need to use energies in our
calculation. But in this framework it is hard to explain the
factor of $\frac{1}{2}$ there and all other calculations in
section \ref{cubic} and section \ref{twopoint2} besides the fact
that we don't know why the free torus two point function should
correspond to wave function renormalization on the string theory
side. Despite lack of interpretations, our calculations
nevertheless set up computational rules to get the right answer
and indeed point out a clear correspondence on both sides. Our
calculations may help to better understand of the question of
holography, which has been address in previous works
\cite{bn,holography,das}. But this question is still unclear. We
do not yet have a clear prescription of what is the correspondence
between the bulk and boundary as we did in the context of AdS/CFT
(by Witten diagram) . Our calculations would provide some sense as
to what specifically do we need to compare on both sides. It would
be interesting to further study this question.

\vspace{0.2in} {\leftline {\bf Acknowledgments}} We thank Vijay
Balasubramanian and  Asad Naqvi for collaborations at early stage
of the project and for readings of the manuscript. We are also
grateful to Thomas S. Levi, Gary Shiu and Matt Strassler for
illuminating discussions.

\appendix

\section{Derivation of symmetry factors}\label{count}
In this appendix we give a practical prescription for deriving
symmetry factors of string theory diagrams we computed. We have
considered two cases of one loop propagation diagrams and one loop
cubic diagrams where the countings are relatively simple (For two
point functions at genus 2 level we would need to count 21 field
theory diagrams \cite{harvmit}) .

\subsection{one loop propagation diagrams}
We denote a close string with n segments by $(a_1a_2\cdots a_n)$,
where the string are regarded as the same by cyclic rotation. For
example, $(a_1a_2\cdots a_n)$ and $(a_2a_3\cdots a_na_1)$ are the
same string. We denote the processes of string splitting and
joining by $(a_1a_2\cdots a_n)\rightarrow(a_1a_2\cdots
a_i)(a_{i+1}\cdots a_n)$ and $(a_1a_2\cdots a_i)(a_{i+1}\cdots
a_n)\rightarrow(a_1a_2\cdots a_n)$. Now imagine figure \ref{F3} as
a string of 4 segments goes from $(1234)$ to $(2143)$. How many
ways can we do this with our rules? A little counting reveal that
at one loop level there are only two processes as the following
\begin{eqnarray}
&&(1234)\rightarrow(12)(34)\rightarrow(2143) \nonumber
\\&& (1234)\rightarrow(23)(41)\rightarrow(2143) \nonumber
\end{eqnarray}
Here since $(12)$ and $(21)$, $(34)$ and (43) are the same, we can
join $(12)(34)$ in to $(2143)$.  These two processes are exactly
one loop string propagation diagrams in figure \ref{F2}. Thus we
conclude we have overcounted by a factor of $2$ when we do string
theory diagram calculations. This explain the symmetry factor of
$\frac{1}{2}$ in equation (\ref{a1}). At this point the meaning of
the procedure may be a little unclear to the readers. The validity
of this procedure will be justified by a less trivial example of
one loop cubic diagrams in the next subsection, in which we find
precise agreements with direct field theory and string theory
diagrams calculation.

\subsection{one loop cubic diagrams}
Now we consider field theory diagrams in figure \ref{F4}. Diagrams
$Q_1$, $Q_2$ and $Q_3$ represent the processes
$(123456)\rightarrow (14325)(6), (36)(1542), (153)(426)$. How many
ways can we go from initial state to final states? For $Q_1$,
there are six processes
\begin{eqnarray}
&&1.~~~(123456)\rightarrow(23)(4561)\rightarrow(325614)\rightarrow(14325)(6)
\nonumber
\\&&
2.~~~(123456)\rightarrow(34)(5612)\rightarrow(432561)\rightarrow(14325)(6)\nonumber
\\&&
3.~~~(123456)\rightarrow(12345)(6)\rightarrow(23)(451)(6)\rightarrow(51432)(6)\nonumber
\\&&
4.~~~(123456)\rightarrow(12345)(6)\rightarrow(34)(512)(6)\rightarrow(43251)(6)\nonumber
\\&&
5.~~~(123456)\rightarrow(34)(5612)\rightarrow(34)(125)(6)\rightarrow(43251)(6)\nonumber
\\&&
6.~~~(123456)\rightarrow(23)(4561)\rightarrow(23)(145)(6)\rightarrow(51432)(6)\nonumber
\end{eqnarray}
As we track the string splitting and joining processes and compare
with string theory diagrams in figure \ref{F5}, we find process
1,2 belong to type $R_2$ string theory diagrams; process 3,4
belong to type $R_1$ string theory diagrams; process 5,6 belong to
type $R_3$ string theory diagrams. For $Q_2$ all possible
processes are
\begin{eqnarray}
&&1.~~~(123456)\rightarrow(234)(156)\rightarrow(423615)\rightarrow(36)(4215)
\nonumber
\\&&
2.~~~(123456)\rightarrow(234)(156)\rightarrow(342156)\rightarrow(36)(4215)\nonumber
\\&&
3.~~~(123456)\rightarrow(12)(3456)\rightarrow(215634)\rightarrow(63)(2154)\nonumber
\\&&
4.~~~(123456)\rightarrow(12)(3456)\rightarrow(12)(45)(36)\rightarrow(36)(2154)\nonumber
\\&&
5.~~~(123456)\rightarrow(45)(1236)\rightarrow(542361)\rightarrow(36)(5421)\nonumber
\\&&
6.~~~(123456)\rightarrow(45)(1236)\rightarrow(45)(12)(36)\rightarrow(36)(5421)\nonumber
\end{eqnarray}
Here process 1,2,3,5 belong to type $R_2$ string theory diagrams;
process 4,6 belong to type $R_3$ string theory diagram. For $Q_3$
all possible processes are
\begin{eqnarray}
&&1.~~~(123456)\rightarrow(123)(456)\rightarrow(312645)\rightarrow(531)(264)
\nonumber
\\&&
2.~~~(123456)\rightarrow(123)(456)\rightarrow(231564)\rightarrow(315)(264)\nonumber
\\&&
3.~~~(123456)\rightarrow(234)(561)\rightarrow(342615)\rightarrow(315)(426)\nonumber
\\&&
4.~~~(123456)\rightarrow(234)(561)\rightarrow(423156)\rightarrow(315)(426)\nonumber
\\&&
5.~~~(123456)\rightarrow(345)(126)\rightarrow(534261)\rightarrow(531)(426)\nonumber
\\&&
6.~~~(123456)\rightarrow(345)(126)\rightarrow(453126)\rightarrow(531)(426)\nonumber
\end{eqnarray}
All processes belong to type $R_2$ string theory diagrams.
Summarizing our results, type $R_1$ diagrams have 2 contributions
from $Q_1$; type $R_2$ diagrams have 2 contributions from $Q_1$, 4
contributions from $Q_2$, 6 contributions from $Q_3$; type $R_3$
diagrams have 2 contributions from $Q_1$, 2 contributions from
$Q_2$. So we conclude
\begin{eqnarray}\label{factor1}
&&\nonumber R_1=2Q_1\\&&\nonumber
R_2=2Q_1+4Q_2+6Q_3\\&&R_3=2Q_1+2Q_2
\end{eqnarray}
This is in exact agreements with equation (\ref{factor}) and
(\ref{factor2}).

\section{Cancellation for chiral/non-chiral amplitude} \label{cancellation}

\begin{figure}
  \begin{center}
 \epsfysize=2.5in
   \mbox{\epsfbox{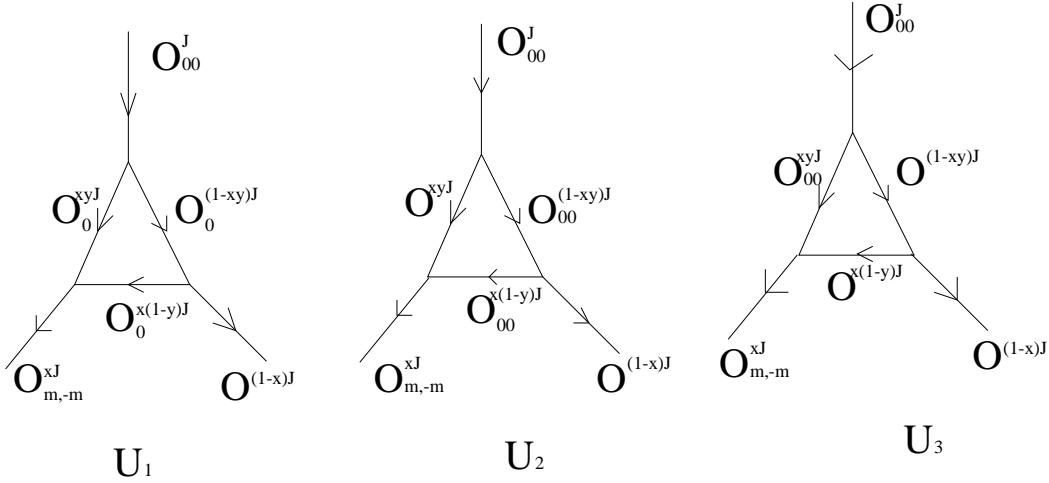}}
\end{center}
\caption{There are three amputated diagrams of
$\langle\bar{O}^J_{00}O^{J_1}_{m,-m}O^{J_2}\rangle_{torus}$. Note
the first diagram $U_1$ has double contribution since there are
two ways to separate the two scalar excitations. Calculations show
the three diagrams exactly cancel each other.} \label{F9}
\end{figure}

From \cite{harvmit} we know the two point function of a chiral
operator with a non-chiral BMN operator vanishes at genus 1 and 2
level. The planar three point functions
$\langle\bar{O}^J_{00}O^{J_1}_{n,-n}O^{J_2}\rangle$ also vanish.
We expect it is generally true that a chiral state can not
propagate or decay into non-chiral states at arbitrary higher
genus. In this appendix we verify
$\langle\bar{O}^J_{00}O^{J}_{n,-n}\rangle_{genus~2}=0$ and
$\langle\bar{O}^J_{00}O^{J_1}_{n,-n}O^{J_2}\rangle_{torus}=0$ from
string theory calculations. Knowing
$\langle\bar{O}^J_{00}O^{J}_{n,-n}\rangle_{torus}=0$ and
$\langle\bar{O}^J_{00}O^{J_1}_{n,-n}O^{J_2}\rangle_{planar}=0$, we
only need to consider amputated diagrams and show they cancel.

\subsection{One loop cubic string interaction}

The diagrams are depicted in figure \ref{F9}. Calculations show
\begin{eqnarray}
2U_1=-2\frac{g_2^3}{\sqrt{J}}x^{\frac{9}{2}}(1-x)^{\frac{1}{2}}\int
_0^1dy~y(1-y)(1-xy)\frac{\sin^2(m\pi y)}{m^2\pi^2}
\end{eqnarray}
\begin{eqnarray}
U_2=U_3=\frac{g_2^3}{\sqrt{J}}x^{\frac{9}{2}}(1-x)^{\frac{1}{2}}\int_0^1
dy~y(1-y)(1-xy)\frac{\sin^2(m\pi y)}{m^2\pi^2}
\end{eqnarray}
Without doing the integral, we can see the contributions of the
three diagrams cancel $2U_1+U_2+U_3=0$.

\begin{figure}
  \begin{center}
 \epsfysize=3.0in
   \mbox{\epsfbox{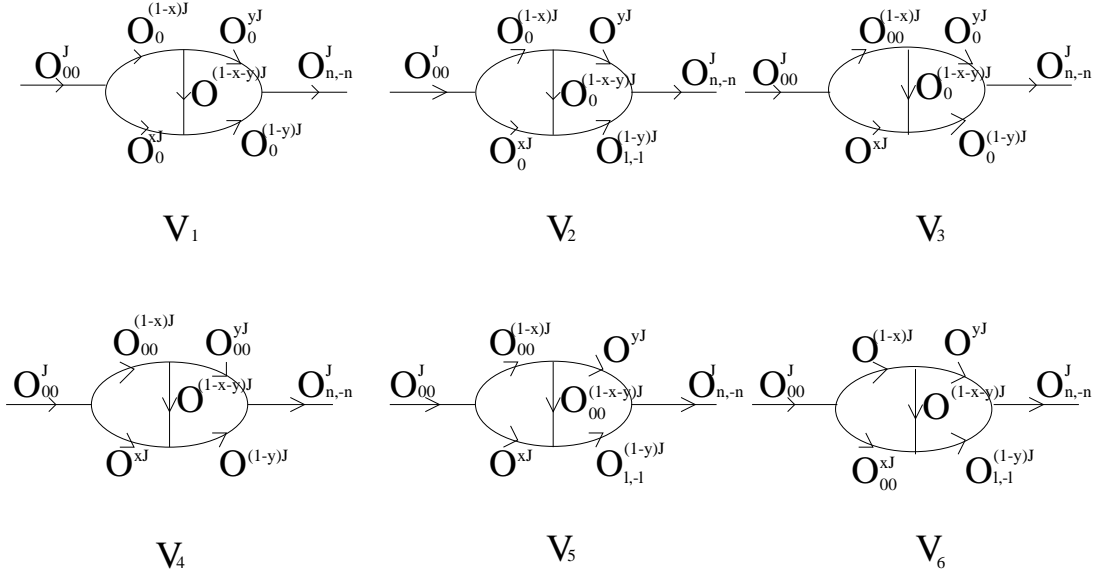}}
\end{center}
\caption{Similar to figure \ref{F8}, there are 6 diagrams
contribute to
$\langle\bar{O}^J_{00}O^{J}_{n,-n}\rangle_{genus~2}$. Again we
note diagrams $V_1$, $V_2$ and $V_3$ have double contributions.}
\label{F10}
\end{figure}
\subsection{Two loop string propagation}

The relevant diagrams are depicted in figure \ref{F10}. As usual
we calculate these diagrams. For example
\begin{eqnarray}
V_1=g_2^4\int_{0<x,y,x+y<1}dxdy~x^2(1-x)y(1-x-y)(-\frac{\sin^2(n\pi
y)}{n^2\pi^2})
\end{eqnarray}
Note diagrams $V_1$, $V_2$ and $V_3$ have double contributions. We
leave the readers to check $2V_1+2V_2+2V_3+V_4+V_5+V_6=0$. The
calculation is simple since the cancellation occurs without doing
the sum and the integral.

\section{Some summation formulae}
\label{summation} We will need to use some useful summation
formulae when we sum over all operators in the string theory loop
diagram. Note the useful identity in \cite{CJ} (see also appendix
D of \cite{GS})

\begin{equation}
\sum_{l=-\infty}^{\infty}(-1)^l
\frac{e^{ily}}{l+\alpha}=\frac{\pi}{\sin(\pi\alpha)}e^{-i\alpha
y}, ~~~~~~~~~~~~~~~~-\pi<y<\pi
\end {equation}
From this equation we can derive (for $0<\beta<1$)
\begin{equation}
\sum_{p\neq0,
p=-\infty}^{\infty}\frac{\sin^2(p\pi\beta)}{(p-\alpha_1)(p-\alpha_2)}=
\frac{\pi}{(\alpha_1-\alpha_2)}[\frac{\sin(\alpha_1\pi(1-\beta))\sin(\alpha_1\pi\beta)}{\sin(\alpha_1\pi)}
-\frac{\sin(\alpha_2\pi(1-\beta))\sin(\alpha_2\pi\beta)}{\sin(\alpha_2\pi)})]
\end{equation}
Then we can take the derivatives of $\alpha_1$ and $\alpha_2$ and
take specific limits of $\alpha_1$ and $\alpha_2$ on both sides of
the equation. Here are some specific identities that will be
useful for our calculations (Assuming $m$ is an integer and
$\alpha$ is not an integer).

\begin{eqnarray}
\sum_{p\neq0}\frac{\sin^2(\beta p\pi)}{(p-\alpha)^2p^2} &=&
-\frac{2\pi}{\alpha
^3}\frac{\sin(\alpha\pi(1-\beta))\sin(\alpha\pi\beta)}{\sin(\alpha\pi)}
\\ \nonumber
&& +
\frac{\pi^2}{\alpha^2}[\frac{(1-\beta)\sin^2(\alpha\pi\beta)+\beta\sin^2
(\alpha\pi(1-\beta))}{\sin^2(\alpha\pi)}+\beta(1-\beta)]
\end{eqnarray}
\begin{eqnarray}
\sum_{p\neq0,p\neq m} \frac{\sin^2(\beta p\pi)}{(p-m)^2p^2} &=&
\frac{1}{6m^4}[-18\sin^2(\beta m \pi)
+(-1+6\beta-6\beta^2)m^2\pi^2\cos(2\beta m\pi) \nonumber
\\&& +(1+6\beta-6\beta^2)m^2\pi^2-6(1-2\beta)m\pi\sin(2\beta
m\pi)]\nonumber
\end{eqnarray}
\begin{eqnarray}
\sum_{p\neq0,p\neq m} \frac{\sin^2(\beta p\pi)}{(p-m)^2p^3} &=&
\frac{1}{6m^5}[-36\sin^2(\beta m \pi)
+(-1+6\beta-6\beta^2)m^2\pi^2\cos(2\beta m\pi) \nonumber
\\&& +(1+12\beta-12\beta^2)m^2\pi^2-9(1-2\beta)m\pi\sin(2\beta
m\pi)]\nonumber
\end{eqnarray}
\begin{eqnarray}
\sum_{p\neq0,p\neq m} \frac{\sin^2(\beta p\pi)}{(p-m)^2p^4} &=&
\frac{1}{6m^6}[-60\sin^2(\beta m \pi)
+(-1+6\beta-6\beta^2))m^2\pi^2\cos(2\beta m\pi) \nonumber
\\&& +(1+18\beta-18\beta^2+2\beta^2(1-\beta)^2m^2\pi^2)m^2\pi^2
\nonumber \\&&-12(1-2\beta)m\pi\sin(2\beta m\pi)]\nonumber
\end{eqnarray}
\begin{eqnarray}
\sum_{p\neq0,p\neq m} \frac{\sin^2(\beta p\pi)}{(p-m)^4p^2} &=&
\frac{1}{90m^6}\{-225+45m^2\pi^2+90\beta(1-\beta)m^2\pi^2+m^4\pi^4
\nonumber \\ && [225-45(1-6\beta+6\beta^2)m^2\pi^2+ \nonumber \\
&& (-1+30\beta^2-60\beta^3+30\beta^4)m^4\pi^4]\cos(2\beta m\pi)
\nonumber \\ && +60(1-2\beta)(-3-\beta m^2\pi^2+\beta^2
m^2\pi^2)m\pi\sin(2\beta m \pi)\} \nonumber
\end{eqnarray}

\end{document}